\documentclass[sigconf]{acmart}
\AtBeginDocument{%
  }

\setcopyright{acmlicensed}
\copyrightyear{2018}
\acmYear{2018}
\acmDOI{XXXXXXX.XXXXXXX}
\acmConference[Conference acronym 'XX]{Make sure to enter the correct
  conference title from your rights confirmation email}{June 03--05,
  2018}{Woodstock, NY}
\acmISBN{978-1-4503-XXXX-X/2018/06}

\newcommand{\eg}{\textit{e.g., }}
\newcommand{\ie}{\textit{i.e., }}
\usepackage{multirow}
\usepackage{tabularray}
\begin{document}

\title{Enhancing COBOL Code Explanations: A Multi-Agents Approach Using Large Language Models}

\author{Fangjian Lei}

\authornotemark[1]
\email{fangjian.lei@queensu.ca}
\affiliation{%
  \institution{Queen's University}
  \city{Kingston}
  \state{Ontario}
  \country{Canada}
}
\author{Jiawen Liu}
\email{jiawen.liu@queensu.ca}
\authornotemark[1]
\affiliation{%
  \institution{Queen's University}
  \city{Kingston}
  \state{Ontario}
  \country{Canada}
}
\author{Shayan Noei}
\email{s.noei@queensu.ca}
\affiliation{%
  \institution{Queen's University}
  \city{Kingston}
  \state{Ontario}
  \country{Canada}
}
\author{Ying Zou}
\email{ying.zou@queensu.ca}
\affiliation{%
  \institution{Queen's University}
  \city{Kingston}
  \state{Ontario}
  \country{Canada}
}
\author{Derek Truong}
\email{trong@ca.ibm.com}
\affiliation{%
  \institution{IBM Canada}
  \city{Markham}
  \state{Ontario}
  \country{Canada}
}
\author{William Alexander}
\email{walexand@us.ibm.com}
\affiliation{%
  \institution{IBM USA}
  \country{United States}
}

\begin{abstract} 
Common Business Oriented Language (COBOL) is a programming language used to develop business applications that are widely adopted by financial, business, and government agencies. Due to its age, complexity, and declining number of COBOL developers, maintaining COBOL codebases is becoming increasingly challenging. In particular, the lack of documentation makes it difficult for new developers to effectively understand and maintain COBOL systems. Existing research utilizes large language models (LLMs) to explain the functionality of code snippets. However, COBOL presents unique challenges due to its architectural and syntactical differences, which often cause its code to exceed the token window size of LLMs. In this work, we propose a multi-agent approach that leverages two LLM-based agents working collaboratively to generate explanations for functions, files, and the overall project. These agents incorporate together by utilizing contextual information from the codebase into the code explanation prompts. We evaluate the effectiveness of our approach using 14 open-source, real-world COBOL projects. Our results indicate that our approach performs significantly better than the baseline in function code explanation, with improvements of 12.67\%, 18.59\%, and 0.62\% in terms of METEOR, chrF, and SentenceBERT scores, respectively.  At the file level, our approach effectively explains both short and long COBOL files that exceed the token window size of LLMs and surpass the baseline by 4.21\%, 10.72\%, and 14.68\% in explaining the purpose, functionality, and clarity of the generated explanation. At the project level, our approach generates explanations that convey the functionality and purpose of 82\% of the selected projects. 
\end{abstract}



\renewcommand\footnotetextcopyrightpermission[1]{}
\setcopyright{none} 
\settopmatter{printacmref=false} 

\maketitle

\section{Introduction}
\label{sec:introduction}
Common Business-Oriented Language (COBOL) is a high-level programming language with a syntax that closely resembles natural English \citep{alfred2007compilers}. COBOL was invented in 1959 to standardize business applications and remains to be widely used in financial institutions, government agencies, and large corporations ~\cite{taulli2020cobol, ciborowska2021contemporary, ali2022cobrex}. Despite its age, COBOL remains crucial to the technological infrastructure of these organizations, powering 80\% of financial service transactions and 95\% of ATM swipes ~\cite{taulli2020cobol}. Moreover, the robustness and reliability of COBOL make it a preferred choice for handling critical tasks, financial transactions, payroll systems, and large-scale database administration ~\cite{taulli2020cobol, ciborowska2021contemporary, ali2022cobrex}. Therefore, many organizations continue to rely on COBOL for its long-term stability and capability to manage complex high-volume processes with consistency and precision. As a result, COBOL has contributed to approximately 220 billion lines of code in use, with 1.5 billion lines written each year as of 2020 ~\cite{taulli2020cobol}. 
The English-like syntax of COBOL offers readability, but often results in longer code for simple operations compared to modern programming languages such as Java and Python~~\cite{sharma2020short}. As reported in prior study ~\cite{kiefer2017cobol}, the average Lines of Code for a COBOL program is 600, while the Java equivalence for the same function is only 30 lines or fewer. Its English-like syntax does not necessarily make it easier to understand and often leads to verbose code. Moreover, legacy COBOL systems often have undergone extensive modifications to accommodate new business needs and hardware updates, without the involvement of the original developers. Therefore, the changes to the codes can obscure its original code structure and intent~~\cite{rajlich1997comprehension}. 
Besides, converting legacy COBOL systems to modern high-level programming language code is a time-consuming, costly, and complex process, as evidenced by the Commonwealth Bank of Australia 5-year \$749.9 million migration project ~\cite{taulli2020cobol}. As a result, to maximize knowledge transfer to newer developers, who require a detailed understanding of system architecture and code~~\cite{de2005study}, it is essential to provide an effective approach to help them understand the legacy COBOL systems.

In recent years, Large Language Models (LLMs) have shown promise in automating the generation of code explanations ~\cite{macneil2022generating,sarsa2022automatic,su2024distilled}. LLMs can effectively analyze and summarize source code and provide detailed explanations of its functionality and structure ~\cite{chen2021evaluating}. Moreover, LLMs can generate code documentation, facilitate software maintenance, and help new developers familiarize themselves with the existing codebase. However, applying LLMs to explain the COBOL code presents two main challenges: (1) The length of COBOL programs often exceeds the LLM's maximum input token limit, preventing the ability to understand the individual source file and the functionality of the entire system. (2) Most research and development efforts have focused on explaining individual functions or small snippets of code ~\cite{leinonen2023comparing,chen2023gptutor,sarsa2022automatic}, but they do not address the demand to understand entire files or systems. 
For legacy COBOL systems, understanding the interactions between different functions and dependencies between files is crucial for effective code maintenance. Therefore, understanding only snippets of code is often inadequate, and comprehensive explanations that encompassing entire systems are essential for effective maintenance, troubleshooting, and refactoring ~\cite{upadhaya2023understanding}.

To provide a complete explanation for a COBOL project, we propose a multi-agents approach that leverages two LLM-based agents to work collaboratively and generate explanation for functions, files, and overall project. This approach incorporates contextual information from the projects into code explanation prompts. More specifically, two types of LLM-based agents include: (1) the \textit{Code Processing Agent}, which analyzes input code snippets and generates corresponding explanations; and (2) the \textit{Text Processing Agent}, which links and expands the explanations generated by the \textit{Code Processing Agent}, merging them into higher-level explanations. By coordinating these agents, we generate COBOL code explanations at the function, file, and project levels. In this study, we conduct our experiments on 14 open-source COBOL projects, aiming to address the following research questions:

\textbf{RQ1: What is the performance of our approach in generating function-level code explanations?}
Accurate function-level code explanations are essential for understanding individual functions, which also builds a foundation for understanding the functionality and purposes of files and the entire project. We generate explanations of function-level COBOL code through the collaboration of the \textit{Code Processing Agent} and the \textit{Text Processing Agent}.
We show that our approach outperforms the baseline with a 12.69\% average improvement in METEOR, a 18.59\% increase in chrF, and a 0.62\% increase in sentenceBERT. By evaluating our approach using \textit{LLM-as-a-Judge Agent}, we confirm that our approach generates more accurate, concise, and informative explanations, achieving an average performance score of 6.01 out of 10, significantly outperforming the baseline performance score of 5.57 out of 10.

\textbf{RQ2: What is the performance of our approach in generating file-level code explanations?}
Generating accurate file-level code explanations is foundational for understanding the overall project, as files often serve as the key units of functionality in COBOL programs. COBOL files are generally long and can exceed the input token limit of the LLMs. Therefore, handling long files requires an effective strategy to bypass the token limit restriction. To address this issue, we employ a hybrid approach to generate the file-level explanation based on file length. For short files, the \textit{Code Processing Agent} generates explanations directly. For long files, we first segment them into functions and then utilize file-level artifacts, such as function dependency relationships, to generate file-level explanations using the \textit{Text Processing Agent}. Our evaluation using \textit{LLM-as-a-Judge Agent} demonstrates that our approach outperforms the baseline across all key evaluation criteria, achieving a 4.21\% improvement in explaining \textit{Purpose} of the file, 10.72\% improvement in explaining \textit{Functionality} of the file, and 14.68\% improvement in \textit{Clarity} of the generated explanations.

\textbf{RQ3: What is the performance of our approach in generating project-level code explanations?}
Project-level explanations are essential for capturing the overall functionality and purpose of COBOL systems. However, due to the token limit size of LLMs, the entire codebase cannot be passed to the LLM as a single prompt. Additionally, understanding the dependency relationships among all the files is required to produce accurate explanations of the projects. To generate a project-level explanation, we leverage project artifacts (\eg file dependencies) and the explanation of each file. The \textit{Text Processing Agent} integrates file-level explanations by accounting for inter-file relationships to generate a project-level summary. Our evaluation using \textit{LLM-as-a-Judge Agent} demonstrates that our approach generates explanations that are as good as or better than 9 out of 11 existing extracted project descriptions.

In this paper, we make the following key contributions:
\vspace{-0.3em}
\begin{itemize}
    \item We propose an approach to generate code explanations for entire COBOL projects using multiple LLM based agents and source code artifacts.
    \item We provide an approach that enables LLMs to explain long files that exceed input token limits by applying a hierarchical merging on the lower-level code explanations.
    \item We provide a dataset that contains open source COBOL code and its corresponding explanations, accessible at: \cite{replicated_package}
\end{itemize}

The remainder of this paper is organized as follows. Section \ref{sec:background} provides an overview of the basic concepts of COBOL. Section \ref{sec:methodology} details the proposed approach. Section \ref{sec:rqs} presents the research questions. Section \ref{sec:implications} discusses the implications. Section \ref{sec:threats_to_validity} addresses the threats to validity. Section \ref{related_work} discusses previous related work. Finally, Section \ref{sec:conclusion} concludes the paper and outlines future work.

\section{COBOL Basic Concepts}\label{sec:background}

\begin{figure}
    \centering
    \includegraphics[width=0.9\linewidth]{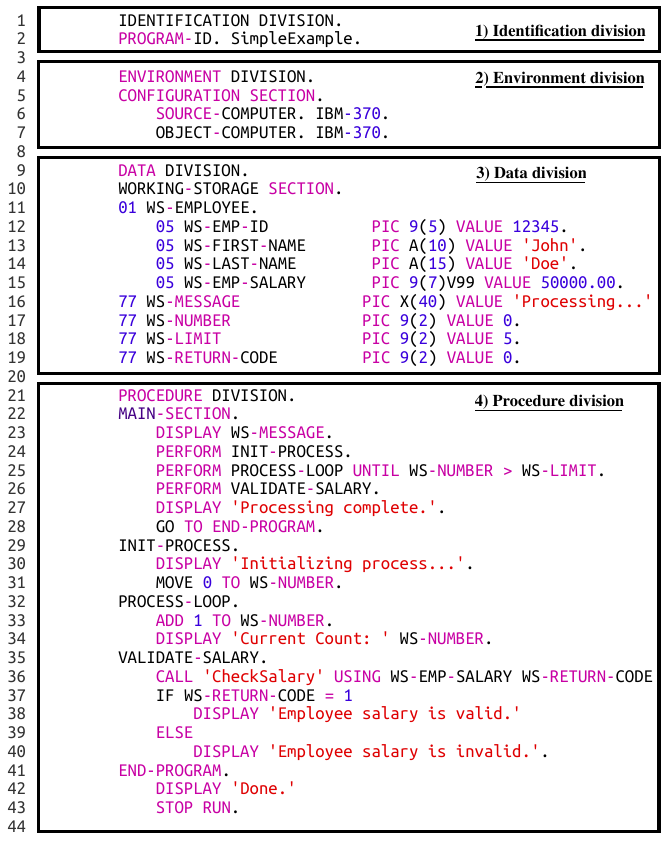}
    \caption{Example of COBOL program that accumulates an employee's salary.}
    \label{fig:cobol_example}
\end{figure}

This section explains the COBOL code and its components. As shown in Figure~\ref{fig:cobol_example}, each COBOL code file consists of four main divisions: (1) Identification, (2) Environment, (3) Data, and (4) Procedure. The purpose of each division is described below:

\begin{enumerate}
    \item \textit{\textbf{Identification division:}} Is the first division of a COBOL file. It defines the file identifier, which is a unique name used to identify the file and is referenced during compilation and execution. A COBOL file can be invoked by other COBOL files using the \texttt{CALL} statement along with its \texttt{PROGRAM-ID}. 

    \item \textit{\textbf{Environment division:}} Specifies the operating environment of a COBOL file, including the configuration of input and output devices.
    
    \item \textit{\textbf{Data division:}} Defines the hierarchy of data structures and variables within a file. In data structure definition, numbers are used to indicate the hierarchical level of the variables, determining their position in the data structure hierarchy.

    \item \textit{\textbf{Procedure Division:}} Defines the executable instructions a file, including the programming logic and operations that manipulate data defined in the \textit{Data Division}. Within the \textit{Procedure Division}, code is organized into paragraphs and sections to create manageable and modular code segments.
    \begin{itemize}
        \item \textbf{Paragraphs}: A paragraph is a named block of code within a section, containing one or more COBOL statements. It can be executed individually using the \texttt{PERFORM} statement, which calls the paragraph by its name. 
        
        For example, as shown in Figure~\ref{fig:cobol_example}, the paragraph \texttt{VALIDATE-SALARY} on line 26 is called. Paragraphs are equivalent to functions in other programming languages, such as Java.
    
        \item \textbf{Sections}: A section is a higher-level organizational unit that groups related paragraphs. For example, in the Figure~\ref{fig:cobol_example}, (lines 22 to 28), \texttt{MAIN-SECTION} sequentially calls paragraphs to complete the salary calculation.
    \end{itemize}
\end{enumerate}
    We refer to a \textit{paragraph} as a \textit{function} in the remainder of this paper.
    
\begin{figure*}[htbp]
    \centering
\includegraphics[width=0.9\textwidth]{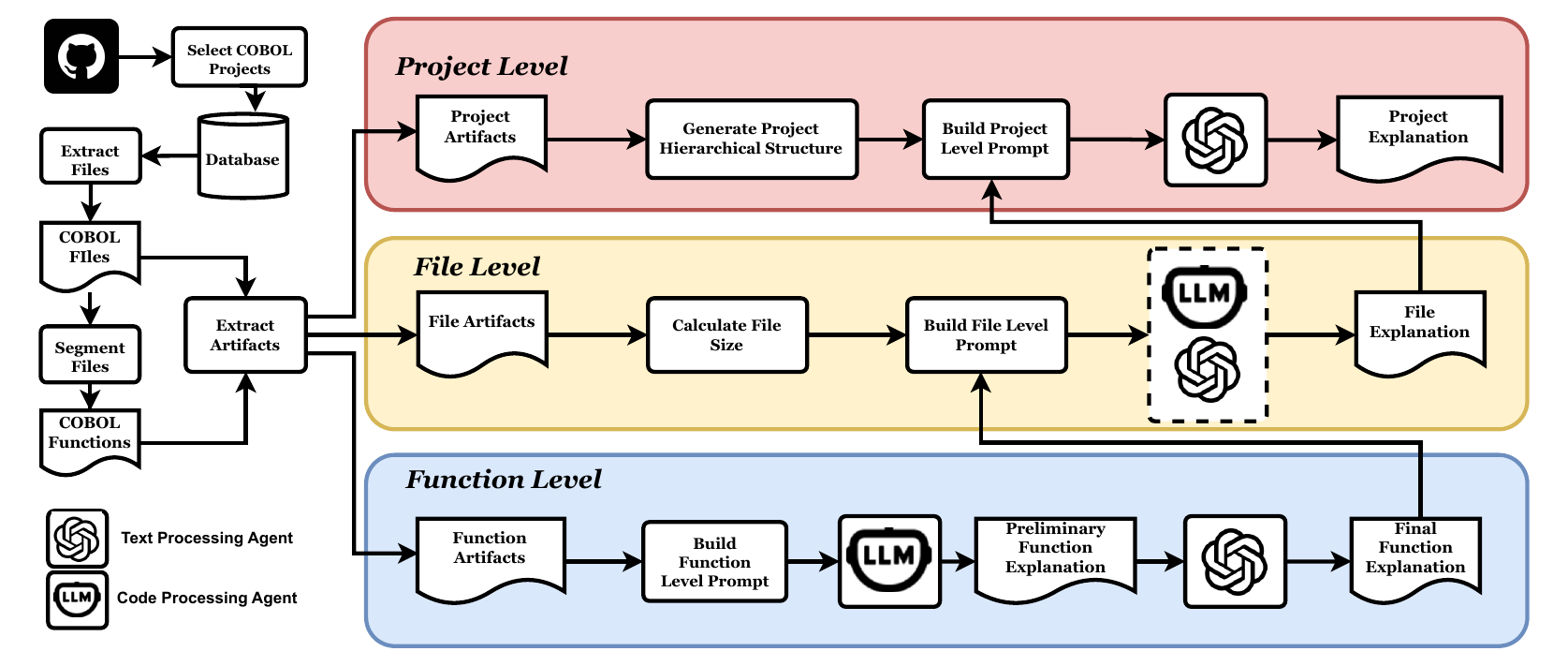}

    \caption{An Overview of Our Multi-agent Approach  on Generating Code Explanation at Function, File and Project Levels.}
    \label{fig:overall_approach}
\end{figure*}



\section{Methodology}
\label{sec:methodology}
In this section, we describe our approach to generate code explanations at different granularities of a COBOL program. 
We select 14 open-source COBOL projects and extract code artifacts at the function, file, and project levels. 
Next, we use two LLM agents, the \textit{Code Processing Agent} and the \textit{Text Processing Agent}, to handle different inputs in our code explanation pipeline (\ie source code and source code explanations).
Lastly, we build three pipelines using our agents for generating code explanations at the function, file, and project levels of a COBOL program.  
An overview of our approach is shown in Figure~\ref{fig:overall_approach}.


\subsection{Project Selection}
\label{sec:PS}
We initially select 84 COBOL projects from a state-of-the-art public COBOL dataset~\cite{ali2023x}. The state-of-the-art COBOL dataset is built by mining GitHub repositories using the GitHub API and filtering out low-quality or irrelevant projects through automated checks and manual review. As IBM\footnote{https://www.ibm.com} is a major vendor of developing and maintaining COBOL applications and the customer information control system (CICS) is one of the most widely used platforms for transactional enterprise COBOL applications\footnote{https://www.ibm.com/products/cics-transaction-server}, we add the two extra COBOL written projects\footnote{https://github.com/cicsdev/cics-banking-sample-application-cbsa}\footnote{https://github.com/cicsdev/cics-genapp} from \textit{IBM CICSdev}\footnote{https://github.com/cicsdev} to our selected projects, thus expanding the subjects of our dataset to 86 projects. With our additions, the final dataset includes all qualified open-source COBOL projects we are able to find.
Moreover, to ensure that the selected projects provide meaningful insights into COBOL code structure and dependencies, we apply the following filtering criteria to refine our selected projects: 
\begin{enumerate}
    \item The project must contain at least two COBOL source files. If a project contains only one file, the project-level explanation is identical to the file-level explanation. 
    \item COBOL must be the primary language, with at least 80\%~\cite{noei2023empirical} of the development code written in COBOL.
\end{enumerate}

After applying the above selection criteria, our final projects consist of 14 open-source projects, including 12 projects from the state-of-the-art dataset ~\cite{ali2023x} and the 2 \textit{IBM CICS} projects. The selected projects are detailed in our replicate package.

\input
\subsection{Reference Dataset Creation}
\label{sec:ground_truth}
We construct a reference dataset to evaluate the code explanations generated by our multi-agent approach at the function, file, and project levels. This dataset acts as a safeguard against hallucinations, ensuring that the generated explanations align with the original intent of the developers.
Below, we detail our reference dataset generation process as follows:
\vspace{-0.3em}
\begin{itemize}
\item \textbf{Function-level reference dataset}: We leverage the functions written by the original developers, which is located immediately above the function names in the source code. In total, we extract 2,316 COBOL functions from our dataset. The first two authors of the paper independently review all the functions, and inspect the correctness of their corresponding comments and evaluate if the comments accurately describe the purpose of the functions. Then, they compare their reviews, and only the comment-function pairs that both annotators agree correctly explain the functions are retained as the function-level reference dataset.

\item \textbf{File-level reference dataset}: We use the header comments provided by the original developers for each file, which describe the functionality and purpose of the entire file, from our dataset. Similar to function-level evaluation, the first two authors independently review 299 files from our dataset and inspect whether the header comments clearly describe the purpose and functionality of the files. Then, the two annotators discuss any conflicts and select the files for which the header comments accurately explain the functionality and purpose of the file. We then refine the selected file-comment pairs and exclude irrelevant content such as copyright notices, author details, revision dates, and non-code-related information.
    \item \textbf{Project-level reference dataset}: We leverage the README files of our projects, as they are the only available documentation that contain the purpose and major functionalities of our projects. The first two authors evaluate each project's README file and select those on which they both agree that the business purpose and overall functionalities are correctly explained.
    We then refine the selected \textit{README} files and remove irrelevant content, such as environment setup or installation instructions, as this information is not relevant to the content in our generated explanations. 
\end{itemize}

Overall, we extract 243 function-comments, 168 file-comments, and 11 project-\textit{README} pairs as our reference dataset from 14 selected open-source projects. The details of the selected COBOL projects, along with our reference datasets, are included in our replication package \cite{replicated_package}.


\subsection{Artifacts Extraction}
\label{sec:artifact_extraction}
To generate documentation at different levels of the project (\ie function, file, and project), we extract different sets of artifacts to provide contextual information for LLMs to better understand the code provided and its dependencies. In the following, we describe the artifacts extracted at each level of granularity:

\subsubsection{Function-Level Artifacts} Explaining a function requires capturing the meaning of its variables and called functions. 
To generate function-level explanations, we extract the following artifacts:

\begin{itemize}
    \item \textbf{Function source code:} is defined after the function name within the \texttt{PROCEDURE DIVISION}.
    \item \textbf{Variable names:}  are declared in the \texttt{DATA DIVISION} before being used in a function.
    \item \textbf{Called functions:} are defined using the \texttt{PERFORM} and \texttt{GO TO} statements.
\end{itemize}

\subsubsection{File-Level Artifacts} Explaining a COBOL file requires understanding its functionalities and structures. We use the following artifacts to generate file-level explanations:

\begin{itemize}
    \item \textbf{Source code:} is defined within \texttt{PROCEDURE DIVISION}
    \item \textbf{Variables:} are declared within \texttt{DATA DIVISION}
    \item \textbf{Function explanations:} are generated using function-level artifacts.
    \item \textbf{Function dependency relationships:} Define each called function within the \textit{Procedure Division }of a file using \texttt{PERFORM} and \texttt{GO TO} statements.
    \item \textbf{File name:}  Specifies the name of the file. 
    \item \textbf{Program ID:} is extracted from the IDENTIFICATION DIVISION.
\end{itemize}

\subsubsection{Project-Level Artifacts} Explaining a COBOL project requires understanding how files connect and interact. File dependencies illustrate the relationships between files and contribute to the overall project explanation. We utilize the following artifacts to generate project-level explanations:
\begin{itemize}
    \item \textbf{File dependency relationships:} Uses the \texttt{LINK} and \texttt{CALL} keywords that a file employs to invoke other files. 
    \item \textbf{Program IDs:} are extracted from each file, including dependent files, from the \texttt{IDENTIFICATION} \texttt{DIVISION}.
    \item \textbf{File explanations:} are generated using file-level artifacts.
\end{itemize}

\noindent We use \textit{NetworkX} \cite{hagberg2008exploring} library to build  dependency graphs based on the extracted file dependencies. In the dependency graph, nodes represent files and edges indicate file calls. We then calculate the topological rankings of the files within the graph, where the files with an in-degree of 0 (\ie they only call other files but are never called) is at the top level and the files with an out-degree of 0 (\ie they are only called but never call other files) is at the bottom level.

\begin{figure*}
    \centering
\includegraphics[width=0.7\textwidth]{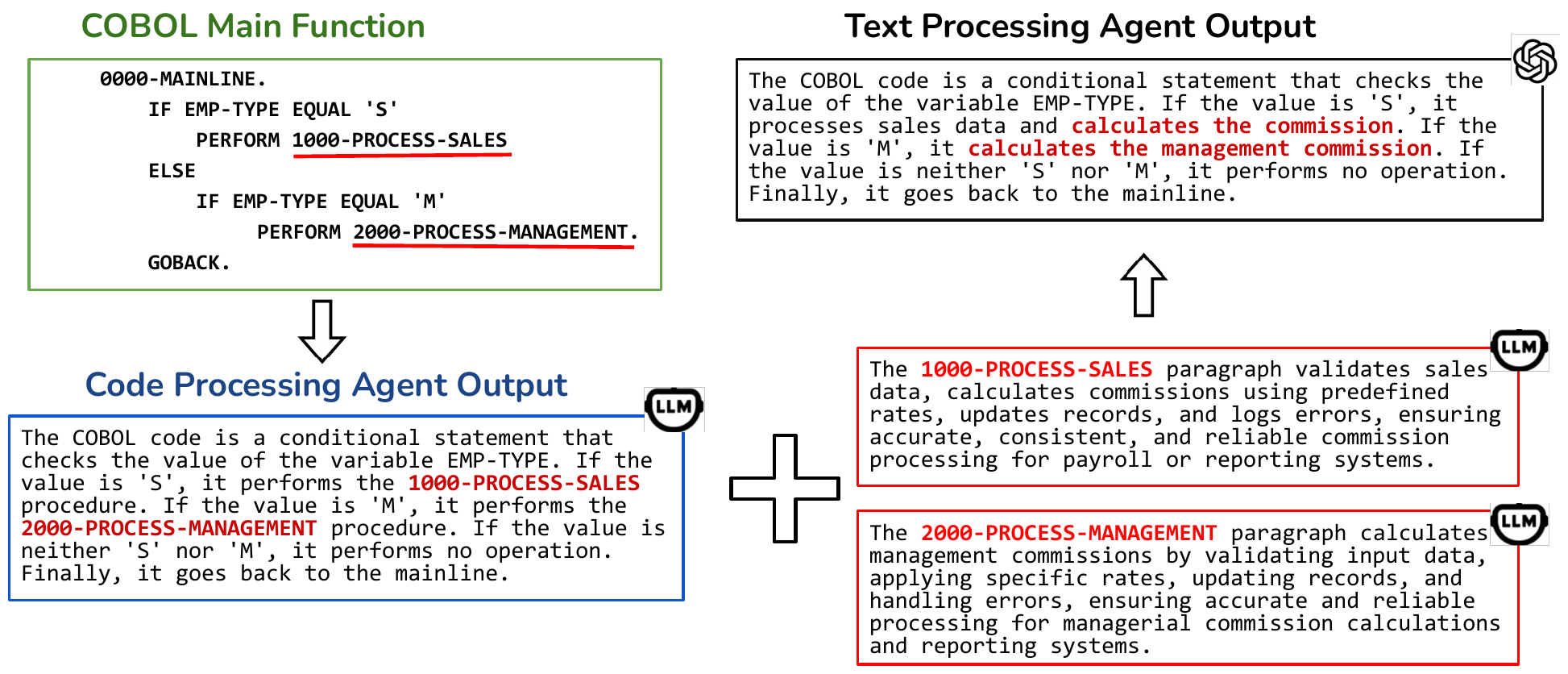}
    \caption{An example of pairing code processing with \textit{Text Processing Agent} in function level explanation}
    \label{fig:motivation_example}
\end{figure*}

\subsection{Agents Selection}
To generate explanations of COBOL code at file, function, and project levels, we use two specialized LLM agents working collaboratively: the \textit{Code Processing Agent} and the \textit{Text Processing Agent}. The \textit{Code Processing Agent} utilizes an LLM that is pre-trained on code from various programming languages, including COBOL, to generate detailed explanations for COBOL code. The \textit{Text Processing Agent}, which is pre-trained on natural language datasets, refines the explanations generated by the \textit{Code Processing Agent} to improve clarity and readability. 
Below are the details of each agent:

\subsubsection{\textbf{Code Processing Agent}}
we use \textit{granite-34b-code-instruct} \cite{mishra2024granite} as our \textit{Code Processing Agent}. \textit{Granite} is an open-source LLM that is trained on COBOL code explanation tasks, making it an effective LLM in understanding and explaining COBOL source codes. \textit{granite-34b-code-instruct} has been recognized as the top performer in enterprise-grade code comprehension \cite{mishra2024granite}. While other models like \textit{XMAiNframe} model family \cite{dau2024xmainframe} are now trained in COBOL,\textit{ granite-34b-code-instruct} was the most advanced publicly available model at the time of our experiments. The \textit{granite-34b-code-instruct} model supports a maximum token size of 8,192 tokens and demonstrates a benchmark accuracy of 76.0\% in code explanation tasks within the \textit{HumanEval} benchmark \cite{chen2021evaluating}.

\subsubsection{\textbf{Text Processing Agent}} \textit{Our Code Processing Agent} is effective at understanding and explaining COBOL code. However, its outputs contain explanations that require further processing to be more readable in natural language. Therefore, we use use \textit{GPT-4o-mini}\footnote{\label{openai}https://platform.openai.com/docs/models} as our \textit{Text Processing Agent} to further refine and merge our \textit{Code Processing Agent} outputs. This model is selected for its speed, cost-effectiveness, and strong performance in generating high-quality, human-readable text. Figure~\ref{fig:motivation_example} shows an example of how the \textit{Text Processing Agent} improves the function-level explanation generated by the \textit{Code Processing Agent}. 
The \textit{Code Processing Agent} generates the explanation of the logic conditions of the COBOL function but does not provide detailed explanations for the called sub-functions, such as \textit{1000-PROCESS-SALES} and \textit{2000-PROCESS-MANAGEMENT}. \textit{The Text Processing Agent} integrates generated explanations of these sub-functions, generated by \textit{the Code Processing Agent}, and merges them into the main function explanation to provide a complete explanation of the function.
 The \textit{GPT-4o-mini} model supports a maximum token size of 128,000 tokens\footref{openai}, allowing it to handle long context effectively. In benchmark evaluations, it can achieve an accuracy of 82.0\% on the Massive Multitask Language Understanding (\textit{MMLU}) benchmark\footnote{https://platform.openai.com/index/gpt-4o-mini-advancing-cost-efficient-intelligence}, which outperforms other small models such as \textit{Gemini Flash} \cite{team2024gemini} and \textit{Claude Haiku} \cite{anthropic2023claude3}. These attributes make \textit{GPT-4o-mini}\footref{openai} an effective choice for transforming technically accurate outputs from our \textit{Code Processing Agent} LLM into final code explanation that aligns with developer expectations.

\subsubsection{\textbf{LLM-as-a-Judge Agent}}
\label{llm_as_a_judge}

We use the \textit{LLM-as-a-Judge Agent} to evaluate our generated explanations because existing text similarity scores focus on text overlap and semantic similarity based on text embeddings, but they do not consider the readability or usefulness of the generated explanations. We use the \textit{LLM-as-a-Judge Agent} to further assess the generated explanations based on informativeness, conciseness, and correctness as suggested by previous studies \cite{fu2023gptscore, liu2023g}. The LLM-as-a-judge assigns a score from 0 to 10 to each explanation with a given criteria, with higher scores indicating better performance. We employ \textit{GPT-4o} as our \textit{LLM-as-a-Judge Agent}, which has been used as a judge in prior studies \cite{zhu2023judgelm,tan2024judgebench} and has demonstrated superior performance in understanding and evaluating complex language tasks \cite{islam2024gpt}. The prompt templates used for our \textit{LLM-as-a-Judge Agent} are included in our replication package.

\subsubsection{\textbf{Agents Environment}}
We utilize an A100 \textit{(NVIDIA)} graphics card with 80 GB of memory to run the \textit{Granite-34b code-instruct} model locally. Furthermore, we use the \textit{OpenAI API}\footnote{\url{https://platform.openai.com/docs/api-reference/introduction}} to access the \textit{GPT-4o mini} and \textit{GPT-4o} models.
\subsection{Explanation Generation}
We integrate our \textit{Text Processing Agent} and \textit{Code Processing Agent} into three distinct pipelines for function, file, and project-level explanation generation. We employ prompting techniques from Chen et. al \cite{chen2023unleashing} and guidelines from the  \textit{OpenAI} cookbook\footnote{\url{https://cookbook.openai.com/articles/related_resources}}, such as output length specifications, to control the verbosity of the responses.
In the following sections, we describe our approach for generating code explanations for each granularity level in detail.

\subsubsection{\textbf{Function-Level Explanation}}
\label{sec:function_explanation}
Functions represent the lowest level of granularity in our study and serve as the input for our file-level explanation generation. Therefore, providing accurate explanations at the function level builds the foundation for generating higher-level code explanations (\ie file and project levels). We generate function-level explanations by following the steps below:

\textbf{Step 1: Generating preliminary function explanations.} For each function in a project, we build the prompt with corresponding function-level artifacts and use the \textit{Code Processing Agent} to generate the preliminary function explanations. 
For each function, we provide the \textit{Code Processing Agent} with a system prompt (included in our replication package), the function code, and the corresponding source code artifacts (\eg called function) as input. We then ask our \textit{Code Processing Agent} to generate the explanation of the target function code snippet. We apply a 75-word threshold to ensure the explanation is concise yet detailed enough to provide the necessary information, as also suggested by OpenAI's prompt length specification technique\footnote{https://platform.openai.com/docs/guides/prompt-engineering}. The 75 words threshold is determined by conducting a gradient search across length thresholds of 50, 75, 100, and 200 words, and finding that 75 words provide the optimal balance between brevity and sufficient details in the generated content. Although the median length of function-level reference dataset is 10 words and the mean is 15 words, we find using a shorter threshold (\eg 50 words or less) often causes the LLM to truncate explanations before completion. Conversely, a higher threshold (\eg over 100 words) tends to produce overly lengthy explanations, sometimes leading to redundant or repeated outputs.

\textbf{Step 2: Generating final function explanations}
The generated function explanation in Step 1 may retain the names of the functions being called without explaining their behavior or their impact on the function that calls them, as illustrated in Figure~\ref{fig:motivation_example}. Moreover, passing the source code of the dependent function to the \textit{Code Processing Agent} can exceed the input size limit. Therefore, if a function calls other functions within the same file, we use the \textit{Text Processing Agent} to refine the explanations generated by the \textit{Code Processing Agent}. Specifically, for functions at the lowest level of the function call hierarchy (\ie functions that do not call any others), the preliminary explanation is the final explanation. For higher-level functions that include calls to other functions in the call hierarchy, we use the \textit{Text Processing Agent} to iteratively refine the preliminary explanations of the called functions. In each iteration, we refine the explanation of any function whose called functions already have final explanations. These refined explanations serve as input for subsequent iterations, and the process continues until all functions have final explanations. 
The prompt template used in this step is available in our replicate package \cite{replicated_package}. 
The system prompt first establishes the role of the LLM as a writing refinement assistant and describes the task. 
In the input prompt, the \textit{\textless{}Main\textgreater{}} tags enclose the explanation of the target function generated by the \textit{Code Processing Agent}, and the \textit{\textless{}Term\textgreater{}} tags enclose the names and explanations of the called functions. 




\subsubsection{\textbf{File-Level Explanation}}
\label{sec:file_explanation}
COBOL files vary in length, with short files having code snippets that fit within the the input size limit of the \textit{Code Processing Agent}, while long files exceed this limit (\ie The \textit{granite-34b-code-instruct} model supports a maximum token size of 8,192 tokens). Therefore, we first calculate the token length of COBOL source code files and classify them as short or long files. Then, we perform the following steps to generate code explanations for short and long files separately.

\textbf{Step 1: Generating short file explanation.} For short files, we prompt the \textit{Code Processing Agent} using file-level artifacts as contextual information to generate file explanations. This approach preserves the original structure and logical flow of the source code within the file and allows the LLM to comprehend the content of the file.
The prompt template used in this step is available in our replicate package \cite{replicated_package}. The system prompt introduces the structure of the input prompt, the task definition, and output specifications. We first provide the program ID and the file name, as we have observed that the program ID and file name often contain information about the business purpose or intended functionality of the file.  Incorporating such details increases the likelihood of including meaningful context for explanation. Next, we put the code of the file under \textit{\textless{}Code\textgreater{}} tags and variables under \textit{\textless{}Variable\textgreater{}} tags.


\textbf{Step 2: Generating long file explanation.} 
Since the file-level artifacts of long files exceed the input limit of the LLM, making direct explanation generation impossible, we employ the hierarchical merging technique \cite{wu2021recursively} that merges explanations of different segments of a file based on hierarchical relationships, allowing us to generate file explanations from them. To this end, we first split the source code into individual functions. Then, using the \textit{Code Explanation Agent}, we generate explanations for the smaller units (\ie functions) and adopt the hierarchical merging prompt template~\cite{chang2023booookscore} to progressively merge the various function-level explanations generated using the \textit{Text Processing Agent}. The prompt template used in this step is available in our replicate package.
We incorporate the \textit{Function Dependency Relationship} between functions into the prompt to ensure the \textit{Text Processing Agent} has an understanding of the function relationship within a COBOL file.
In this system prompt, we define the task of recursively merging function explanations based on function call dependency and introduce the input prompt structure. For the input prompt, we include \textit{Program Id}, \textit{File name}, and function names along with their explanations. Additionally, we include the \textit{Function Dependency Relationship} enclosed by the \textit{\textless{}Relationship\textgreater{}} tags.

\begin{table*}
\centering
\caption{Text similarity results for function-level explanations comparing the generated results with the reference dataset. The Signif. Diff row reports the statistical significance (based on p-values) and the effect size measured using Cliff’s delta.}
\label{tab:paragraph_automatic_result}
\begin{tabular}{l|ll|ll|ll} 
\hline
\multirow{2}{*}{\begin{tabular}[c]{@{}l@{}}\textbf{Evaluation }\textbf{ Metrics}\end{tabular}} & \multicolumn{2}{l|}{\textbf{METEOR}}   & \multicolumn{2}{l|}{\textbf{ChrF}}      & \multicolumn{2}{l}{\textbf{SentenceBERT}}  \\
                                                                                                 & \textbf{Mean}       & \textbf{Median}   & \textbf{Mean}       & \textbf{Median}    & \textbf{Mean}      & \textbf{Median}        \\ 
\hline
\textbf{Baseline}                                                                                & 0.134              & 0.126             & 0.199              & 0.202              & 0.4047            & 0.4095                 \\ 
\hline
\textbf{Our Approach}                                                                            & 0.151              & 0.135             & 0.236              & 0.242              & 0.4072            & 0.4101                 \\ 
\hline
\textbf{Improvement}                                                                             & 12.69\% $\uparrow$ & 7.14\% $\uparrow$ & 18.59\% $\uparrow$ & 19.80\% $\uparrow$ & 0.62\% $\uparrow$ & 0.15\% $\uparrow$      \\ 
\hline
\textbf{Signif. Diff}                                                                            & \multicolumn{2}{l|}{* (negligible)}    & \multicolumn{2}{l|}{*** (small)}        & \multicolumn{2}{l}{not   significant}      \\ 
\hline
\multicolumn{7}{l}{Signif. codes: p \textbf{0 ‘***’ 0.001 ‘**’ 0.01 ‘*’ 0.05} ‘.’ 0.1 ‘ ’ 1}                                                                                                                                     \\
\hline
\end{tabular}
\end{table*}


\subsubsection{\textbf{Project-Level Explanation}}
\label{sec:project_explanation} Similar to generating explanations for long files, the entire source code of a project cannot fit within the input limit of the LLM. Therefore, we employ hierarchical merging to integrate individual file-level explanations into a project-level explanation. This is achieved by incorporating project level artifacts (\eg file dependency relationships) into the project-level prompt, ensuring that all relationships and interactions between files are preserved.
We build the project-level prompt for the \textit{Text Processing Agent} to generate project explanations. 
The prompt template used in this step is available in our replicate package \cite{replicated_package}. In the project-level prompt template, we include the file names and their corresponding explanations for all files in the project and provide \textit{File Relationship Dependencies} (\ie topological rankings of the files) within the \textless{}Structures\textgreater{} tags.
\section{Research Question Results} 
\label{sec:rqs}
In this section, we provide the motivation, approach, and results of our research questions.\\

\noindent\textbf{RQ1: What is the performance of our approach in generating function-level code explanations?}
\subsubsection*{\textbf{Motivation}}
Function-level code explanations can significantly enhance code clarity and long-term maintainability\cite{garousi2015usage}; therefore, they are an essential part of the code documentation process. Furthermore, as explained in Section~\ref{sec:file_explanation}, functions serve as the foundational components for generating file-level explanations in our approach. Consequently, any errors in function-level explanations would be propagated to file-level and project-level explanations. In this research question, we aim to evaluate the effectiveness of our approach in generating function-level explanations.

\subsubsection*{\textbf{Approach}}
As explained in Section~\ref{sec:function_explanation}, we incorporate function-level artifacts into our \textit{Code Processing Agent} prompt. Then, we use a \textit{Text Processing Agent} to refine and merge the outputs of the \textit{Code Processing Agent}, generating function-level explanations. To evaluate our generated explanations, we first establish a baseline approach that uses zero-shot prompting, where the \textit{Code Processing Agent} is asked to directly explain the function without providing additional function-level artifacts. We then compare our generated function-level code explanations with baseline explanations. \\
\noindent\textbf{Alignment evaluation of the generated explanations:}
We first use three text similarity metrics to assess the alignment of the generated function-level explanations with the reference dataset (\ie function comments). The rationale is that a high similarity to the original comment indicates that our generated function-level explanations are not hallucinated and remain closely aligned with the original documentation. The details of the similarity metrics are explained below:
\begin{itemize}
    \item \textit{METEOR}~\cite{banerjee2005meteor}: Measures the similarity between the generated text and the reference text based on precision, recall, and alignment of unigrams, with additional emphasis on synonyms, stemming, and paraphrasing.
    \item \textit{ChrF}~\cite{popovic2015chrf}: Calculates the similarity between texts based on character n-grams, providing a more fine-grained assessment of linguistic similarity.
    \item \textit{SentenceBERT}~\cite{reimers2019sentence}: Uses a sentence-level embedding model to calculate semantic similarity between generated explanations and reference comments, capturing deeper contextual meaning.
\end{itemize}

\noindent\textbf{Evaluation of generated explanations against baseline:} To evaluate the explanations generated by our approach, we utilize the \textit{LLM-as-a-judge Agent} (described in Section~\ref{llm_as_a_judge}), which compares the explanations produced by our approach and the baseline, using the reference dataset (e.g, function comments) as a benchmark. This helps determine which explanation aligns more closely with the original intent.
Our function-level \textit{LLM-as-a-judge} prompt is adapted from the pairwise evaluation prompt proposed by Zhu et al. \cite{zhu2023judgelm}. To prevent bias towards either explanations, we do not disclose which explanation corresponds to our approach or the baseline. The function-level judge prompt is included in our replication package. The evaluation is based on the following three criteria:
\begin{enumerate}
    \item Clarity: COBOL-specific terms should not appear directly unless properly explained, ensuring the explanation is more understandable;
    \item Conciseness: Explanations should be clear and concise;
    \item Correctness: Explanations must accurately describe the function's behavior.
\end{enumerate}


\noindent\textbf{Manual Verification:}
We conduct a manual evaluation to further verify the reliability and consistency of the evaluations performed by our \textit{LLM-as-a-judge Agent}. To this end, with a confidence level of 90\% and a 10\% margin of error \cite{junk1999confidence}, we randomly sample 54 functions from our function-level reference dataset. Our manual evaluation assesses two aspects: (1) \textit{reference coverage} and (2) \textit{hallucination detection}. \textit{Reference coverage} measures how much the information describing the function-level reference dataset is included in the generated explanation. \textit{Hallucination detection} evaluates whether the explanation correctly describes the COBOL function and does not hallucinate. Two annotators (\ie the first two authors) independently label all sampled functions and assign a score on a 0 to 1 scale for each function explanation, where:
\begin{itemize}
    \item 0 indicates no relevant information;
    \item 0.25 represents less than half of the expected content;
    \item 0.5 indicates roughly half is covered;
    \item 0.75 reflects more than half;
    \item 1 denotes a fully correct explanation.
\end{itemize}
\indent For example, if the reference states that a function performs four operations, and the generated explanation correctly describes three, the reference coverage score would be 0.75. The Cohen’s kappa \cite{cohen1960coefficient} scores are 0.65 for reference coverage and 0.58 for hallucination detection, indicating substantial and moderate agreements, respectively. Any conflicts are then discussed and resolved to reach a final agreement.

\subsubsection*{\textbf{Results:}}

\noindent\textbf{Our approach effectively generates COBOL function explanations that closely align with developer-provided comments, outperforming the baseline in METEOR and chrF.} 
Table~\ref{tab:paragraph_automatic_result} presents the average and median results of our text similarity measurement evaluation for function-level explanation generation. We observe that the similarity of our generated explanations at the function level is 0.41, indicating a moderate \cite{bhattacharya2022legal} similarity between our explanations and the reference explanations (\ie function comments). This suggests that our generated results align with the original intent of the developers. Furthermore, our approach outperforms the baseline in all text similarity metrics, METEOR, chrF, and sentenceBERT, with average improvements of 12.69\%, 18.59\%, and 0.62\%, respectively. We further use a two-sided Mann-Whitney U test~\cite{nachar2008mann} to statistically compare the differences in the text similarity scores between our approach and the baseline. The test results confirm that improvements in METEOR and chrF scores are significant (\textit{p-value=0.01 and 0.00}), with an effect size of negligible and small respectively.  

\begin{table}
\centering
\caption{LLM-as-a-Judge evaluation results for function-level. The Signif. Diff row reports the statistical significance (based on p-values) and the effect size measured using Cliff’s delta.}
\label{tab:llm_judge_results_function_level}
\begin{tabular}{l|ll|ll} 
\hline
                       & \multicolumn{2}{l|}{\textbf{Baseline}} & \multicolumn{2}{l}{\textbf{Our Approach}}  \\ 
\hline
                       & \textbf{Mean} & \textbf{Median}         & \textbf{Mean} & \textbf{Median}             \\ 
\hline
\textbf{Overall Score} & 5.57         & 5.0                     & 6.01         & 6.0                         \\ 
\hline
\textbf{Signif. Diff}  & \multicolumn{4}{l}{***(small)}                                                      \\ 
\hline
\multicolumn{5}{l}{Signif. codes: p \textbf{0 ‘***’ 0.001 ‘**’ 0.01 ‘*’ 0.05} ‘.’ 0.1 ‘ ’ 1}                 \\
\hline
\end{tabular}
\end{table}

\noindent\textbf{Our approach effectively generates COBOL function explanations that are more clear, concise, and accurate than the baseline.} Table \ref{tab:llm_judge_results_function_level} presents the evaluation results for function-level COBOL explanations, assessed using our \textit{LLM-as-a-Judge Agent}. To compare our approach with the baseline, we perform a two-sided Mann-Whitney U test and compare the differences in the overall score of the \textit{LLM-as-a-Judge Agent} between our approach and the baseline. Our approach demonstrates significantly better performance (\textit{p-value=0.001}) compared to the baseline, with an average score of 6.01 and a median score of 6.0. Therefore, our approach produces function-level explanations that are more informative, concise, and accurate compared to the baseline. 

\noindent\textbf{We verify that our approach generates accurate code descriptions and does not impose hallucinations, through manual evaluation.}
Our manual evaluation results shows that the median reference coverage and hallucination detection scores are both 1 with the means of 0.773 and 0.995 respectively. These results indicate that the generated descriptions closely align with the original COBOL code comments, with no hallucinations observed. 

\noindent\textbf{Providing code artifacts enhances the performance of LLMs in generating code explanations at the function level.} Our approach integrates code artifacts, such as variable names, into the \textit{Code Processing Agent} and uses the \textit{Text Processing Agent} with the called functions and their corresponding explanations to generate complete function explanations. As a result, the generated explanations are better aligned with the original developer comments compared to the baseline, as demonstrated by both the LLM-as-a-Judge Agent and our manual verification results.\\

\noindent\textbf{RQ2:What is the performance of our approach in generating file-level code explanations?}
\subsubsection*{\textbf{Motivation}}
File-level explanations are essential for understanding the overall functionality and flow of a program~\cite{aghajani2019software}, especially for new developers who need an overview before exploring function-level details \cite{de2005study}. Accurate file-level explanations are essential for generating project-level explanations in our approach. This research question aims to assess the reliability of our file-level code explanation generation and determine whether incorporating file-level artifacts leads to better code explanations compared to the baseline.

\begin{table}
\centering
\caption{\textit{LLM-as-a-judge} evaluation results for file-level. The Signif. Diff row reports the statistical significance (based on p-values) and the effect size measured using Cliff’s delta.}
\label{tab:llm_judge_results_file_level}
\begin{tabular}{l|l|l|l} 
\hline
\multirow{2}{*}{\begin{tabular}[c]{@{}l@{}}\textbf{Evaluation }\\\textbf{ Metrics}\end{tabular}} & \textbf{Purpose}  & \textbf{Functionality} & \textbf{Clarity}    \\
                                                                                                 & \textbf{(Mean)}   & \textbf{(Mean)}        & \textbf{(Mean)}     \\ 
\hline
\textbf{Baseline}                                                                                & 6.41              & 5.69                   & 5.79                \\ 
\hline
\textbf{Our Approach}                                                                            & 6.68              & 6.30                   & 6.64                \\ 
\hline
\textbf{Improvement}                                                                             & 4.21\% $\uparrow$ & 10.72\% $\uparrow$     & 14.68\% $\uparrow$  \\ 
\hline
\textbf{Signif. Diff}                                                                            & *** (small)       & *** (small)            & **** (medium)       \\
\hline
\end{tabular}
\end{table}

\subsubsection*{\textbf{Approach}} 
To measure the quality of file-level explanations, we use a zero-shot prompting approach as our baseline. Then, using our reference dataset, we evaluate our generated code explanations with the baseline.

\noindent\textbf{Evaluation of generated file explanations against baseline:} 
We use our \textit{LLM-as-a-Judge Agent} to compare the file-level explanations with the baseline while checking their alignment with the reference dataset. The \textit{LLM-as-a-Judge Agent} uses the following criteria to compare our code explanations with the baseline: 
\vspace{-0.3em}
\begin{itemize}
    \item Purpose: The explanation accurately conveys the intent and potential usage of the file, aligning with the reference comment.
    \item Functionality: The explanation correctly describes the functionality of the file as stated in the reference file description.
    \item Clarity: The explanation is well-structured, easy to follow, and free from ambiguity.
\end{itemize}

\noindent\textbf{Manual evaluation:}
Due to the input token limit of our \textit{Code Processing Agent}, \textit{LLM-as-a-Judge Agent} is used to compare our approach and the baseline only on short files. Thus, we evaluate the long files through manual assessment, which also covers both short and long files. Our manual evaluation also helps to verify the reliability of our \textit{LLM-as-a-Judge Agent}. Similar to the evaluation metrics used for our \textit{LLM-as-a-Judge Agent}, we employ \textit{Purpose}, \textit{Functionality}, and \textit{Completeness}. \textit{Completeness} is similar to \textit{Clarity}, but with the added consideration of how well the sub-functions within the main function are clearly explained, which \textit{LLM-as-a-Judge Agent} is not capable of assessing due to its limited input size.
We follow the same manual evaluation procedure as in RQ1 and randomly select 22 long and 50 short file-comment pairs. The two annotators achieve Cohen’s kappa \cite{cohen1960coefficient} scores of 0.78 for Purpose, 0.58 for Functionality, and 0.72 for Completeness. These results indicate substantial agreement for \textit{Purpose} and \textit{Completeness}, and moderate agreement for \textit{Functionality}. All conflicts are then discussed and resolved.

\subsubsection*{\textbf{Results}}

\textbf{Our approach effectively generates file-level COBOL code explanations, surpassing the baseline in \textit{Purpose} (4.21\%), \textit{Functionality} (10.72\%), and \textit{Clarity} (14.68\%).} 
Table \ref{tab:llm_judge_results_file_level} presents the results of the \textit{LLM-as-a-Judge Agent} evaluation results for file-level explanations. Our approach provides better explanations by improving 4.21\% in \textit{Purpose}, a 10.72\% in \textit{Functionality}, and a 14.68\% in \textit{Clarity} over the baseline. The results of the two-sided Mann-Whitney U test (\textit{p-values $<$ 0.05}) show that our approach significantly outperforms the baseline in providing clear purpose and functionality for the file explanations.



\textbf{Our approach can effectively generate the COBOL code explanation at the file level for both long and short files}. The results of our manual investigation, shown in Table~\ref{tab:file_manual_result}, confirms that the explanations generated by our approach align closely with the reference dataset. Specifically, our approach achieves a mean score of 0.85, 1.00, and 0.93 for short files, and 0.85, 1.00, and 0.97 for long files in Purpose, Functionality, and Completeness, respectively.
The results of the manual investigation further validate our LLM-as-a-judge findings and suggest that our approach can serve as a robust solution for automated COBOL file explanation generation, effectively capturing both the file’s intent and its functional details.\\

\begin{table}[th]
\centering
\caption{Manual Evaluation Result for File-Level Explanations Compared to the Reference Dataset.}
\label{tab:file_manual_result}
\begin{tabular}{l|l|l|l} 
\hline
\begin{tabular}[c]{@{}l@{}}\textbf{Evaluation }\\\textbf{Metrics}\end{tabular} & \begin{tabular}[c]{@{}l@{}}\textbf{ Purpose}\\\textbf{(Mean)}\end{tabular} & \begin{tabular}[c]{@{}l@{}}\textbf{Functionality}\\\textbf{(Mean)}\end{tabular} & \begin{tabular}[c]{@{}l@{}}\textbf{Completeness}\\\textbf{(Mean)}\end{tabular}  \\ 
\hline
\textbf{Long Files}                                                            & 0.85                                                                       & 0.99                                                                            & 0.91                                                                            \\ 
\hline
\textbf{Short Files}                                                           & 0.84                                                                       & 1.00                                                                            & 0.97                                                                            \\ 
\hline
\textbf{Overall }                                                              & 0.85                                                                       & 1.00                                                                            & 0.93                                                                            \\
\hline
\end{tabular}
\end{table}

\noindent\textbf{RQ3: What is the performance of our approach in generating project-level code explanations?}
\subsubsection*{\textbf{Motivation}}
Project-level explanations, such as the project summary in the README file, provide developers and users with a quick overview of a project. However, README files are not always available \cite{venigalla2022empirical,prana2021including} or sufficiently informative, as they may not cover major functionalities and usage instructions. Hence, comprehensive documentation on the overall system is useful for developers to understand it efficiently and reduce the time required to comprehend it solely by reading the source code. In this research question, we aim to generate project explanations that include file dependency information and assess the effectiveness of incorporating project-level artifacts in producing more informative explanations.
\subsubsection*{\textbf{Approach}}
To evaluate the performance of project-level explanations, we employ the \textit{LLM-as-a-Judge Agent} and perform a manual evaluation to determine whether the generated explanations provide a clearer and more informative project overview compared to the reference dataset (\ie README file).

\textbf{Evaluation generated of project explanations.} We use our \textit{LLM-as-a-Judge Agent} to evaluate the generated project explanations against the reference dataset. The \textit{LLM-as-a-Judge Agent} is presented with the reference data (\ie README) and explanation generated by our approach and asked to determine which description provides a better explanation of the COBOL project and assess whether the two descriptions are conceptually similar, indicating whether they convey comparable project explanation. We included our project-level judge prompt in our replication package.  The rationale behind asking conceptually is that our goal is to generate project-level explanations for COBOL projects without existing README files. If our generated project-level explanation is similar to the README, it can potentially serve as an effective alternative to README documentation.

\textbf{Manual evaluation.} To validate the reliability of the \textit{LLM-as-a-judge} evaluation, we conduct a manual evaluation comparing the content of our generated explanations to the corresponding README files, following the same procedure as in RQ1 and RQ2. Specifically, two annotators manually label if the generated explanations contain the information described in the corresponding reference dataset README documentation, and assign a score of 1 for a successful match. The Cohen’s kappa score \cite{cohen1960coefficient} for the labeling is 0.85, indicating near-perfect agreement. Any discrepancies are discussed and resolved.

\subsubsection*{\textbf{Results}}
\textbf{Our approach successfully generates project descriptions that match or surpass the README files in 9 out of 11 tested COBOL projects.} 
Based on the \textit{LLM-as-a-Judge Agent} results, our approach successfully generates project explanations that match or surpass the README files in 9 out of 11 COBOL projects, indicating its effectiveness in capturing project details. 

\textbf{Our manual evaluation results confirm that 10 out of 11 cases 
successfully explain the functionalities of the project.} Our manual evaluation mostly verifies the \textit{LLM-as-a-Judge Agent} results and shows the consistency between LLM evaluation and human evaluation. For \textit{health-apis} project, the LLM-judge assess that the README provides a better explanation, while our manual evaluation indicates the opposite. We find this discrepancy is caused by unclear descriptions in the README file which could influence the LLM's judgment. For example, the README provides a broad and abstract description: \textit{“This code pattern shows you how to expose data from a Db2 database and a CICS application through z/OS Connect Enterprise Edition and then create APIs to access that data with API Connect....”}
 However, our generated explanation is more concrete and detailed, explicitly specifying the types of data involved: \textit{“The COBOL system comprises a structured hierarchy of files designed to manage patient data and medication processes within a CICS environment....interaction with patient data, medication management, and HTTP request handling....such as blood pressure and heart rate data, while others handle the addition of medications, patient records, and user interactions....”}
Our approach provides more details, particularly by identifying specific data types and functionalities, which are not explicitly stated in the README. This additional detail is from file-level explanations, where we analyze the structure of COBOL files and extract variable types to infer the actual data being processed. This suggests that the LLM judge’s evaluation may be influenced by the abstraction level of the README and could be biased towards more abstract descriptions.
For the one project (\ie \textit{zECS} project) where both the \textit{LLM-as-a-Judge} Agent and our manual evaluation confirm that the README file provides a better explanation, may be due to the broader system's usage and purpose which cannot be determined only from the code.
Therefore, the README describes this project as a cloud-enabled distributed caching service, but our generated explanation does not explicitly capture this concept. We find that all COBOL files within this project handle web requests, but the broader usage of the \textit{zECS} system as a caching service is not directly evident from the code itself. 

\textbf{Project-level explanations generated by our approach effectively encapsulate the purpose for all projects.} Through manual investigation, we find that although the generated explanations may not fully align with the README description, they convey a similar purpose. Therefore, our approach can generate project explanations comparable to README documentation in most cases, and as a result, it has the potential to automatically generate README files. Our approach offers a reliable alternative for COBOL projects that either lack a README or have outdated documentation.

\section{Implications}
\label{sec:implications}
In this section, we provide the implications of our study for researchers and developers.

\noindent\textbf{Enhancing code comprehension for new developers.} By providing explanations at different granularities of a COBOL project, our approach facilitates knowledge transfer and reduces reliance on previous potential out-of-date documentation. Therefore, new COBOL developers can contribute more effectively without the need for extensive experience or direct knowledge transfer from the original developers.

\noindent\textbf{Bridging the gap in legacy code documentation.} While modern programming languages like Java and Python benefit from extensive documentation tools, COBOL systems remain under-supported despite their critical role in business operations. Our work helps developers and managers overcome this limitation and maintain code documentation for their COBOL codebases more effectively.

\noindent\textbf{Overcoming Token Size Limits in LLM-Based Source Code Analysis.} By segmenting long programs into smaller units and using the hierarchical merging technique, we bypass the token size limit of the LLMs. Researchers, developers, and tool-makers can adopt our methodology when constrained by token limits in code analysis. While some current LLMs may offer longer context windows, COBOL codebases are also growing in size and complexity. Our proposed approach will remain essential for producing scalable and accurate documentation. More importantly, directly feeding entire oversized COBOL files into a model is neither practical nor effective — as file sizes increase, brute-force approaches quickly face technical and performance limitations.

\noindent\textbf{Providing a Generalizable Code Comprehension Approach.} LLMs are capable of understanding code in a wide range of programming languages. The core techniques of our approach, including artifact extraction, prompt composition, code slicing, and hierarchical merging, can be applied to other languages supported by LLMs. Consequently, developers working in languages beyond COBOL can also benefit from enhanced code understanding using our method.


\section{Threats to validity}
\label{sec:threats_to_validity}
In this section, we discuss the threats to the validity of our study.

\textbf{Threats to construct validity} concern the data preparation and feature selection. Our dataset is predominantly made up of IBM projects (\ie 7 out of 14). Five projects are extracted from a public COBOL dataset, with two additional projects manually added, as discussed in Section~\ref{sec:PS}. This dominance may introduce bias, as the characteristics and structures of these projects might not fully represent the various COBOL applications. Therefore, our findings might be influenced by the specific coding practices and conventions found in these projects, potentially limiting the generalizability of our results. However, since IBM is a major vendor of COBOL code for mainframe servers, our work remains highly relevant and applicable to a significant number of COBOL developers.

\textbf{Threats to internal validity} define the validity of the methods used in our approach. We build and test our approach using \textit{Granite-34b-code-instruct} and \textit{GPT-4-mini}. The \textit{Granite-34b-code-instruct} model is selected because it is open source and demonstrates good performance in COBOL code comprehension, while the \textit{GPT-4-mini} model is chosen for its good performance and cost-efficiency in text generation. Although the performance of our approach may vary with other LLMs, we believe that the our approach provides a robust pipeline for code explanation generation.

\textbf{Threats to external validity} concern the generalizability of our findings. 
Our approach is focused on generating documentation from COBOL code, which has distinct code structures, programming styles, and syntactical features that may not be as same as other programming languages. Therefore, extending our approach to other programming languages that follow different programming paradigms (\eg inheritance) may require adjustments to our workflow to consider the unique characteristics of these languages.

\section{Related Work}
\label{sec:related_work}
This section reviews the literature on code explanation generation.
\label{related_work}

\textbf{Static Code Comprehension.} Prior work has explored code summarization and comment generation, especially for languages like Java. For example, Steidl et al.\cite{6613836} propose a taxonomy of code comments, Moreno et al.\cite{6613830} develop Java class summaries using lexical tools, and Burden and Heldal~\cite{burden2011natural} generate documentation from class diagrams. Sridhara et al.\cite{sridhara2010towards, sridhara2011automatically, sridhara2011generating} extract keywords and high-level actions for method summaries, while McBurney et al.\cite{mcburney2014automatic,7203110} improve summaries by incorporating context and analyzing method calls. Java also benefits from various documentation tools~\cite{hewage2024automatic,zhai2016automatic}. In contrast, COBOL lacks such tool support, motivating our use of pre-trained LLM agents to generate multi-level code explanations (function, file, and project).

\textbf{LLM approaches to code explanation.} 
Ahmed et al.~\cite{ahmed2024automatic} introduce a few-shot prompt template that contains contextual information extracted from the source code to generate summaries for code snippets. Weisong Sun et al.~\cite{sun2024source} evaluate LLMs for code summarization, focusing on evaluation methods, prompting, model configurations, and cross-language performance. Toufique Ahmed et al.~\cite{ahmed2024automatic} enhance code summarization by augmenting prompts with semantic facts extracted from source code, improving performance across settings. However, due to the token window size limitations of the LLMs, the current approach is restricted to summarizing code snippets. Our approach overcomes the limitations in the existing work and allows for generating more comprehensive summaries at function, file, and project levels.

\vspace{-1em}
\section{Conclusion and Future Work}\label{sec:conclusion}
COBOL systems are still widely used in financial and business agencies. However, there are limited approaches for generating documentation for COBOL compared to languages such as Java. Furthermore, existing LLM-based code explanation generation primarily suffers from the input token limits of LLMs. To overcome the existing limitations, we leverage two specialized agents: a \textit{Code Processing Agent} and a \textit{Text Processing Agent}, which utilize source code artifacts, such as function dependencies, and hierarchical merging of their outputs to provide code explanations at different granularities, including the function, file, and project levels.
Our approach outperforms the baseline methods in generating function, file, and project-level explanations in terms of clarity, conciseness, correctness, and completeness. Our approach reduces the reliance of new developers on existing out-of-date documentation by generating code explanations using the available source code. In the future, we plan to extend our approach to other programming languages.
\vspace{-1em}
\section{Reproducibility}
The replication package of the study can be accessed at: \cite{replicated_package}

\bibliographystyle{ACM-Reference-Format}
\bibliography{main}


\begin{thebibliography}{54}


\ifx \showCODEN    \undefined \def \showCODEN     #1{\unskip}     \fi
\ifx \showISBNx    \undefined \def \showISBNx     #1{\unskip}     \fi
\ifx \showISBNxiii \undefined \def \showISBNxiii  #1{\unskip}     \fi
\ifx \showISSN     \undefined \def \showISSN      #1{\unskip}     \fi
\ifx \showLCCN     \undefined \def \showLCCN      #1{\unskip}     \fi
\ifx \shownote     \undefined \def \shownote      #1{#1}          \fi
\ifx \showarticletitle \undefined \def \showarticletitle #1{#1}   \fi
\ifx \showURL      \undefined \def \showURL       {\relax}        \fi
\providecommand\bibfield[2]{#2}
\providecommand\bibinfo[2]{#2}
\providecommand\natexlab[1]{#1}
\providecommand\showeprint[2][]{arXiv:#2}

\bibitem[Aghajani et~al\mbox{.}(2019)]%
        {aghajani2019software}
\bibfield{author}{\bibinfo{person}{Emad Aghajani}, \bibinfo{person}{Csaba Nagy}, \bibinfo{person}{Olga~Lucero Vega-M{\'a}rquez}, \bibinfo{person}{Mario Linares-V{\'a}squez}, \bibinfo{person}{Laura Moreno}, \bibinfo{person}{Gabriele Bavota}, {and} \bibinfo{person}{Michele Lanza}.} \bibinfo{year}{2019}\natexlab{}.
\newblock \showarticletitle{Software documentation issues unveiled}. In \bibinfo{booktitle}{\emph{2019 IEEE/ACM 41st International Conference on Software Engineering (ICSE)}}. IEEE, \bibinfo{pages}{1199--1210}.
\newblock


\bibitem[Ahmed et~al\mbox{.}(2024)]%
        {ahmed2024automatic}
\bibfield{author}{\bibinfo{person}{Toufique Ahmed}, \bibinfo{person}{Kunal~Suresh Pai}, \bibinfo{person}{Premkumar Devanbu}, {and} \bibinfo{person}{Earl Barr}.} \bibinfo{year}{2024}\natexlab{}.
\newblock \showarticletitle{Automatic semantic augmentation of language model prompts (for code summarization)}. In \bibinfo{booktitle}{\emph{Proceedings of the IEEE/ACM 46th International Conference on Software Engineering}}. \bibinfo{pages}{1--13}.
\newblock


\bibitem[Alfred et~al\mbox{.}(2007)]%
        {alfred2007compilers}
\bibfield{author}{\bibinfo{person}{V~Aho Alfred}, \bibinfo{person}{S~Lam Monica}, {and} \bibinfo{person}{D~Ullman Jeffrey}.} \bibinfo{year}{2007}\natexlab{}.
\newblock \bibinfo{booktitle}{\emph{Compilers principles, techniques \& tools}}.
\newblock \bibinfo{publisher}{pearson Education}.
\newblock


\bibitem[Ali et~al\mbox{.}(2022)]%
        {ali2022cobrex}
\bibfield{author}{\bibinfo{person}{Mir~Sameed Ali}, \bibinfo{person}{Nikhil Manjunath}, {and} \bibinfo{person}{Sridhar Chimalakonda}.} \bibinfo{year}{2022}\natexlab{}.
\newblock \showarticletitle{COBREX: A Tool for Extracting Business Rules from COBOL}. In \bibinfo{booktitle}{\emph{2022 IEEE International Conference on Software Maintenance and Evolution (ICSME)}}. IEEE, \bibinfo{pages}{464--468}.
\newblock


\bibitem[Ali et~al\mbox{.}(2023)]%
        {ali2023x}
\bibfield{author}{\bibinfo{person}{Mir~Sameed Ali}, \bibinfo{person}{Nikhil Manjunath}, {and} \bibinfo{person}{Sridhar Chimalakonda}.} \bibinfo{year}{2023}\natexlab{}.
\newblock \showarticletitle{X-cobol: A dataset of cobol repositories}.
\newblock \bibinfo{journal}{\emph{arXiv preprint arXiv:2306.04892}} (\bibinfo{year}{2023}).
\newblock


\bibitem[Anonymous(2025)]%
        {replicated_package}
\bibfield{author}{\bibinfo{person}{Anonymous}.} \bibinfo{year}{2025}\natexlab{}.
\newblock \bibinfo{booktitle}{\emph{Enhancing COBOL Code Explanations: A Multi-Agents Approach Using Large Language Models}}.
\newblock
\urldef\tempurl%
\url{https://github.com/anonymous-987654321/ICSE2026}
\showURL{%
\tempurl}
\newblock
\shownote{Accessed: 2025-03-14}.


\bibitem[Anthropic(2023)]%
        {anthropic2023claude3}
\bibfield{author}{\bibinfo{person}{Anthropic}.} \bibinfo{year}{2023}\natexlab{}.
\newblock \bibinfo{title}{The Claude 3 Model Family: Opus, Sonnet, Haiku}.
\newblock
\urldef\tempurl%
\url{https://www-cdn.anthropic.com/de8ba9b01c9ab7cbabf5c33b80b7bbc618857627/Model_Card_Claude_3.pdf}
\showURL{%
\tempurl}
\newblock
\shownote{Accessed: 18 May 2024}.


\bibitem[Banerjee and Lavie(2005)]%
        {banerjee2005meteor}
\bibfield{author}{\bibinfo{person}{Satanjeev Banerjee} {and} \bibinfo{person}{Alon Lavie}.} \bibinfo{year}{2005}\natexlab{}.
\newblock \showarticletitle{METEOR: An automatic metric for MT evaluation with improved correlation with human judgments}. In \bibinfo{booktitle}{\emph{Proceedings of the acl workshop on intrinsic and extrinsic evaluation measures for machine translation and/or summarization}}. \bibinfo{pages}{65--72}.
\newblock


\bibitem[Bhattacharya et~al\mbox{.}(2022)]%
        {bhattacharya2022legal}
\bibfield{author}{\bibinfo{person}{Paheli Bhattacharya}, \bibinfo{person}{Kripabandhu Ghosh}, \bibinfo{person}{Arindam Pal}, {and} \bibinfo{person}{Saptarshi Ghosh}.} \bibinfo{year}{2022}\natexlab{}.
\newblock \showarticletitle{Legal case document similarity: You need both network and text}.
\newblock \bibinfo{journal}{\emph{Information Processing \& Management}} \bibinfo{volume}{59}, \bibinfo{number}{6} (\bibinfo{year}{2022}), \bibinfo{pages}{103069}.
\newblock


\bibitem[Burden and Heldal(2011)]%
        {burden2011natural}
\bibfield{author}{\bibinfo{person}{H{\aa}kan Burden} {and} \bibinfo{person}{Rogardt Heldal}.} \bibinfo{year}{2011}\natexlab{}.
\newblock \showarticletitle{Natural language generation from class diagrams}. In \bibinfo{booktitle}{\emph{Proceedings of the 8th International Workshop on Model-Driven Engineering, Verification and Validation}}. \bibinfo{pages}{1--8}.
\newblock


\bibitem[Chang et~al\mbox{.}(2023)]%
        {chang2023booookscore}
\bibfield{author}{\bibinfo{person}{Yapei Chang}, \bibinfo{person}{Kyle Lo}, \bibinfo{person}{Tanya Goyal}, {and} \bibinfo{person}{Mohit Iyyer}.} \bibinfo{year}{2023}\natexlab{}.
\newblock \showarticletitle{Booookscore: A systematic exploration of book-length summarization in the era of llms}.
\newblock \bibinfo{journal}{\emph{arXiv preprint arXiv:2310.00785}} (\bibinfo{year}{2023}).
\newblock


\bibitem[Chen et~al\mbox{.}(2023b)]%
        {chen2023unleashing}
\bibfield{author}{\bibinfo{person}{Banghao Chen}, \bibinfo{person}{Zhaofeng Zhang}, \bibinfo{person}{Nicolas Langren{\'e}}, {and} \bibinfo{person}{Shengxin Zhu}.} \bibinfo{year}{2023}\natexlab{b}.
\newblock \showarticletitle{Unleashing the potential of prompt engineering in Large Language Models: a comprehensive review}.
\newblock \bibinfo{journal}{\emph{arXiv preprint arXiv:2310.14735}} (\bibinfo{year}{2023}).
\newblock


\bibitem[Chen et~al\mbox{.}(2023a)]%
        {chen2023gptutor}
\bibfield{author}{\bibinfo{person}{Eason Chen}, \bibinfo{person}{Ray Huang}, \bibinfo{person}{Han-Shin Chen}, \bibinfo{person}{Yuen-Hsien Tseng}, {and} \bibinfo{person}{Liang-Yi Li}.} \bibinfo{year}{2023}\natexlab{a}.
\newblock \showarticletitle{GPTutor: a ChatGPT-powered programming tool for code explanation}. In \bibinfo{booktitle}{\emph{International Conference on Artificial Intelligence in Education}}. Springer, \bibinfo{pages}{321--327}.
\newblock


\bibitem[Chen et~al\mbox{.}(2021)]%
        {chen2021evaluating}
\bibfield{author}{\bibinfo{person}{Mark Chen}, \bibinfo{person}{Jerry Tworek}, \bibinfo{person}{Heewoo Jun}, \bibinfo{person}{Qiming Yuan}, \bibinfo{person}{Henrique Ponde De~Oliveira Pinto}, \bibinfo{person}{Jared Kaplan}, \bibinfo{person}{Harri Edwards}, \bibinfo{person}{Yuri Burda}, \bibinfo{person}{Nicholas Joseph}, \bibinfo{person}{Greg Brockman}, {et~al\mbox{.}}} \bibinfo{year}{2021}\natexlab{}.
\newblock \showarticletitle{Evaluating large language models trained on code}.
\newblock \bibinfo{journal}{\emph{arXiv preprint arXiv:2107.03374}} (\bibinfo{year}{2021}).
\newblock


\bibitem[Ciborowska et~al\mbox{.}(2021)]%
        {ciborowska2021contemporary}
\bibfield{author}{\bibinfo{person}{Agnieszka Ciborowska}, \bibinfo{person}{Aleksandar Chakarov}, {and} \bibinfo{person}{Rahul Pandita}.} \bibinfo{year}{2021}\natexlab{}.
\newblock \showarticletitle{Contemporary COBOL: Developers' perspectives on defects and defect location}. In \bibinfo{booktitle}{\emph{2021 IEEE International Conference on Software Maintenance and Evolution (ICSME)}}. IEEE, \bibinfo{pages}{227--238}.
\newblock


\bibitem[Cohen(1960)]%
        {cohen1960coefficient}
\bibfield{author}{\bibinfo{person}{Jacob Cohen}.} \bibinfo{year}{1960}\natexlab{}.
\newblock \showarticletitle{A coefficient of agreement for nominal scales}.
\newblock \bibinfo{journal}{\emph{Educational and psychological measurement}} \bibinfo{volume}{20}, \bibinfo{number}{1} (\bibinfo{year}{1960}), \bibinfo{pages}{37--46}.
\newblock


\bibitem[Dau et~al\mbox{.}(2024)]%
        {dau2024xmainframe}
\bibfield{author}{\bibinfo{person}{Anh~TV Dau}, \bibinfo{person}{Hieu~Trung Dao}, \bibinfo{person}{Anh~Tuan Nguyen}, \bibinfo{person}{Hieu~Trung Tran}, \bibinfo{person}{Phong~X Nguyen}, {and} \bibinfo{person}{Nghi~DQ Bui}.} \bibinfo{year}{2024}\natexlab{}.
\newblock \showarticletitle{XMainframe: A Large Language Model for Mainframe Modernization}.
\newblock \bibinfo{journal}{\emph{arXiv preprint arXiv:2408.04660}} (\bibinfo{year}{2024}).
\newblock


\bibitem[de~Souza et~al\mbox{.}(2005)]%
        {de2005study}
\bibfield{author}{\bibinfo{person}{Sergio Cozzetti~B de Souza}, \bibinfo{person}{Nicolas Anquetil}, {and} \bibinfo{person}{K{\'a}thia~M de Oliveira}.} \bibinfo{year}{2005}\natexlab{}.
\newblock \showarticletitle{A study of the documentation essential to software maintenance}. In \bibinfo{booktitle}{\emph{Proceedings of the 23rd annual international conference on Design of communication: documenting \& designing for pervasive information}}. \bibinfo{pages}{68--75}.
\newblock


\bibitem[Fu et~al\mbox{.}(2023)]%
        {fu2023gptscore}
\bibfield{author}{\bibinfo{person}{Jinlan Fu}, \bibinfo{person}{See-Kiong Ng}, \bibinfo{person}{Zhengbao Jiang}, {and} \bibinfo{person}{Pengfei Liu}.} \bibinfo{year}{2023}\natexlab{}.
\newblock \showarticletitle{Gptscore: Evaluate as you desire}.
\newblock \bibinfo{journal}{\emph{arXiv preprint arXiv:2302.04166}} (\bibinfo{year}{2023}).
\newblock


\bibitem[Garousi et~al\mbox{.}(2015)]%
        {garousi2015usage}
\bibfield{author}{\bibinfo{person}{Golara Garousi}, \bibinfo{person}{Vahid Garousi-Yusifo{\u{g}}lu}, \bibinfo{person}{Guenther Ruhe}, \bibinfo{person}{Junji Zhi}, \bibinfo{person}{Mahmoud Moussavi}, {and} \bibinfo{person}{Brian Smith}.} \bibinfo{year}{2015}\natexlab{}.
\newblock \showarticletitle{Usage and usefulness of technical software documentation: An industrial case study}.
\newblock \bibinfo{journal}{\emph{Information and software technology}}  \bibinfo{volume}{57} (\bibinfo{year}{2015}), \bibinfo{pages}{664--682}.
\newblock


\bibitem[Hagberg et~al\mbox{.}(2008)]%
        {hagberg2008exploring}
\bibfield{author}{\bibinfo{person}{Aric Hagberg}, \bibinfo{person}{Pieter~J Swart}, {and} \bibinfo{person}{Daniel~A Schult}.} \bibinfo{year}{2008}\natexlab{}.
\newblock \bibinfo{booktitle}{\emph{Exploring network structure, dynamics, and function using NetworkX}}.
\newblock \bibinfo{type}{{T}echnical {R}eport}. \bibinfo{institution}{Los Alamos National Laboratory (LANL), Los Alamos, NM (United States)}.
\newblock


\bibitem[Hewage(2024)]%
        {hewage2024automatic}
\bibfield{author}{\bibinfo{person}{Nipuni~T Hewage}.} \bibinfo{year}{2024}\natexlab{}.
\newblock \showarticletitle{Automatic Migration of Java Platform Threads to Virtual Threads}.
\newblock  (\bibinfo{year}{2024}).
\newblock


\bibitem[Islam and Moushi(2024)]%
        {islam2024gpt}
\bibfield{author}{\bibinfo{person}{Raisa Islam} {and} \bibinfo{person}{Owana~Marzia Moushi}.} \bibinfo{year}{2024}\natexlab{}.
\newblock \showarticletitle{Gpt-4o: The cutting-edge advancement in multimodal llm}.
\newblock \bibinfo{journal}{\emph{Authorea Preprints}} (\bibinfo{year}{2024}).
\newblock


\bibitem[Junk(1999)]%
        {junk1999confidence}
\bibfield{author}{\bibinfo{person}{Thomas Junk}.} \bibinfo{year}{1999}\natexlab{}.
\newblock \showarticletitle{Confidence level computation for combining searches with small statistics}.
\newblock \bibinfo{journal}{\emph{Nuclear Instruments and Methods in Physics Research Section A: Accelerators, Spectrometers, Detectors and Associated Equipment}} \bibinfo{volume}{434}, \bibinfo{number}{2-3} (\bibinfo{year}{1999}), \bibinfo{pages}{435--443}.
\newblock


\bibitem[Kiefer(2017)]%
        {kiefer2017cobol}
\bibfield{author}{\bibinfo{person}{Charles Kiefer}.} \bibinfo{year}{2017}\natexlab{}.
\newblock \bibinfo{title}{COBOL as a modern language}.
\newblock


\bibitem[Leinonen et~al\mbox{.}(2023)]%
        {leinonen2023comparing}
\bibfield{author}{\bibinfo{person}{Juho Leinonen}, \bibinfo{person}{Paul Denny}, \bibinfo{person}{Stephen MacNeil}, \bibinfo{person}{Sami Sarsa}, \bibinfo{person}{Seth Bernstein}, \bibinfo{person}{Joanne Kim}, \bibinfo{person}{Andrew Tran}, {and} \bibinfo{person}{Arto Hellas}.} \bibinfo{year}{2023}\natexlab{}.
\newblock \showarticletitle{Comparing code explanations created by students and large language models}. In \bibinfo{booktitle}{\emph{Proceedings of the 2023 Conference on Innovation and Technology in Computer Science Education V. 1}}. \bibinfo{pages}{124--130}.
\newblock


\bibitem[Liu et~al\mbox{.}(2023)]%
        {liu2023g}
\bibfield{author}{\bibinfo{person}{Yang Liu}, \bibinfo{person}{Dan Iter}, \bibinfo{person}{Yichong Xu}, \bibinfo{person}{Shuohang Wang}, \bibinfo{person}{Ruochen Xu}, {and} \bibinfo{person}{Chenguang Zhu}.} \bibinfo{year}{2023}\natexlab{}.
\newblock \showarticletitle{G-eval: Nlg evaluation using gpt-4 with better human alignment}.
\newblock \bibinfo{journal}{\emph{arXiv preprint arXiv:2303.16634}} (\bibinfo{year}{2023}).
\newblock


\bibitem[MacNeil et~al\mbox{.}(2022)]%
        {macneil2022generating}
\bibfield{author}{\bibinfo{person}{Stephen MacNeil}, \bibinfo{person}{Andrew Tran}, \bibinfo{person}{Dan Mogil}, \bibinfo{person}{Seth Bernstein}, \bibinfo{person}{Erin Ross}, {and} \bibinfo{person}{Ziheng Huang}.} \bibinfo{year}{2022}\natexlab{}.
\newblock \showarticletitle{Generating diverse code explanations using the gpt-3 large language model}. In \bibinfo{booktitle}{\emph{Proceedings of the 2022 ACM Conference on International Computing Education Research-Volume 2}}. \bibinfo{pages}{37--39}.
\newblock


\bibitem[McBurney(2015)]%
        {7203110}
\bibfield{author}{\bibinfo{person}{Paul~W. McBurney}.} \bibinfo{year}{2015}\natexlab{}.
\newblock \showarticletitle{Automatic Documentation Generation via Source Code Summarization}. In \bibinfo{booktitle}{\emph{2015 IEEE/ACM 37th IEEE International Conference on Software Engineering}}, Vol.~\bibinfo{volume}{2}. \bibinfo{pages}{903--906}.
\newblock
\href{https://doi.org/10.1109/ICSE.2015.288}{doi:\nolinkurl{10.1109/ICSE.2015.288}}


\bibitem[McBurney and McMillan(2014)]%
        {mcburney2014automatic}
\bibfield{author}{\bibinfo{person}{Paul~W McBurney} {and} \bibinfo{person}{Collin McMillan}.} \bibinfo{year}{2014}\natexlab{}.
\newblock \showarticletitle{Automatic documentation generation via source code summarization of method context}. In \bibinfo{booktitle}{\emph{Proceedings of the 22nd International Conference on Program Comprehension}}. \bibinfo{pages}{279--290}.
\newblock


\bibitem[Mishra et~al\mbox{.}(2024)]%
        {mishra2024granite}
\bibfield{author}{\bibinfo{person}{Mayank Mishra}, \bibinfo{person}{Matt Stallone}, \bibinfo{person}{Gaoyuan Zhang}, \bibinfo{person}{Yikang Shen}, \bibinfo{person}{Aditya Prasad}, \bibinfo{person}{Adriana~Meza Soria}, \bibinfo{person}{Michele Merler}, \bibinfo{person}{Parameswaran Selvam}, \bibinfo{person}{Saptha Surendran}, \bibinfo{person}{Shivdeep Singh}, {et~al\mbox{.}}} \bibinfo{year}{2024}\natexlab{}.
\newblock \showarticletitle{Granite code models: A family of open foundation models for code intelligence}.
\newblock \bibinfo{journal}{\emph{arXiv preprint arXiv:2405.04324}} (\bibinfo{year}{2024}).
\newblock


\bibitem[Moreno et~al\mbox{.}(2013)]%
        {6613830}
\bibfield{author}{\bibinfo{person}{Laura Moreno}, \bibinfo{person}{Jairo Aponte}, \bibinfo{person}{Giriprasad Sridhara}, \bibinfo{person}{Andrian Marcus}, \bibinfo{person}{Lori Pollock}, {and} \bibinfo{person}{K. Vijay-Shanker}.} \bibinfo{year}{2013}\natexlab{}.
\newblock \showarticletitle{Automatic generation of natural language summaries for Java classes}. In \bibinfo{booktitle}{\emph{2013 21st International Conference on Program Comprehension (ICPC)}}. \bibinfo{pages}{23--32}.
\newblock
\href{https://doi.org/10.1109/ICPC.2013.6613830}{doi:\nolinkurl{10.1109/ICPC.2013.6613830}}


\bibitem[Nachar et~al\mbox{.}(2008)]%
        {nachar2008mann}
\bibfield{author}{\bibinfo{person}{Nadim Nachar} {et~al\mbox{.}}} \bibinfo{year}{2008}\natexlab{}.
\newblock \showarticletitle{The Mann-Whitney U: A test for assessing whether two independent samples come from the same distribution}.
\newblock \bibinfo{journal}{\emph{Tutorials in quantitative Methods for Psychology}} \bibinfo{volume}{4}, \bibinfo{number}{1} (\bibinfo{year}{2008}), \bibinfo{pages}{13--20}.
\newblock


\bibitem[Noei et~al\mbox{.}(2023)]%
        {noei2023empirical}
\bibfield{author}{\bibinfo{person}{Shayan Noei}, \bibinfo{person}{Heng Li}, \bibinfo{person}{Stefanos Georgiou}, {and} \bibinfo{person}{Ying Zou}.} \bibinfo{year}{2023}\natexlab{}.
\newblock \showarticletitle{An Empirical Study of Refactoring Rhythms and Tactics in the Software Development Process}.
\newblock \bibinfo{journal}{\emph{IEEE Transactions on Software Engineering}} \bibinfo{volume}{49}, \bibinfo{number}{12} (\bibinfo{year}{2023}), \bibinfo{pages}{5103--5119}.
\newblock


\bibitem[Popovi{\'c}(2015)]%
        {popovic2015chrf}
\bibfield{author}{\bibinfo{person}{Maja Popovi{\'c}}.} \bibinfo{year}{2015}\natexlab{}.
\newblock \showarticletitle{chrF: character n-gram F-score for automatic MT evaluation}. In \bibinfo{booktitle}{\emph{Proceedings of the tenth workshop on statistical machine translation}}. \bibinfo{pages}{392--395}.
\newblock


\bibitem[Prana et~al\mbox{.}(2021)]%
        {prana2021including}
\bibfield{author}{\bibinfo{person}{Gede Artha~Azriadi Prana}, \bibinfo{person}{Denae Ford}, \bibinfo{person}{Ayushi Rastogi}, \bibinfo{person}{David Lo}, \bibinfo{person}{Rahul Purandare}, {and} \bibinfo{person}{Nachiappan Nagappan}.} \bibinfo{year}{2021}\natexlab{}.
\newblock \showarticletitle{Including everyone, everywhere: Understanding opportunities and challenges of geographic gender-inclusion in oss}.
\newblock \bibinfo{journal}{\emph{IEEE Transactions on Software Engineering}} \bibinfo{volume}{48}, \bibinfo{number}{9} (\bibinfo{year}{2021}), \bibinfo{pages}{3394--3409}.
\newblock


\bibitem[Rajlich(1997)]%
        {rajlich1997comprehension}
\bibfield{author}{\bibinfo{person}{Vaclav Rajlich}.} \bibinfo{year}{1997}\natexlab{}.
\newblock \showarticletitle{Comprehension and evolution of legacy software (tutorial)}. In \bibinfo{booktitle}{\emph{Proceedings of the 19th international conference on Software engineering}}. \bibinfo{pages}{669--670}.
\newblock


\bibitem[Reimers(2019)]%
        {reimers2019sentence}
\bibfield{author}{\bibinfo{person}{N Reimers}.} \bibinfo{year}{2019}\natexlab{}.
\newblock \showarticletitle{Sentence-BERT: Sentence Embeddings using Siamese BERT-Networks}.
\newblock \bibinfo{journal}{\emph{arXiv preprint arXiv:1908.10084}} (\bibinfo{year}{2019}).
\newblock


\bibitem[Sarsa et~al\mbox{.}(2022)]%
        {sarsa2022automatic}
\bibfield{author}{\bibinfo{person}{Sami Sarsa}, \bibinfo{person}{Paul Denny}, \bibinfo{person}{Arto Hellas}, {and} \bibinfo{person}{Juho Leinonen}.} \bibinfo{year}{2022}\natexlab{}.
\newblock \showarticletitle{Automatic generation of programming exercises and code explanations using large language models}. In \bibinfo{booktitle}{\emph{Proceedings of the 2022 ACM Conference on International Computing Education Research-Volume 1}}. \bibinfo{pages}{27--43}.
\newblock


\bibitem[Sharma(2020)]%
        {sharma2020short}
\bibfield{author}{\bibinfo{person}{Mamillapally~Raghavender Sharma}.} \bibinfo{year}{2020}\natexlab{}.
\newblock \showarticletitle{A short communication on computer programming languages in modern era}.
\newblock \bibinfo{journal}{\emph{Int. J. Comput. Sci. Mob. Comput}}  \bibinfo{volume}{9} (\bibinfo{year}{2020}), \bibinfo{pages}{50--60}.
\newblock


\bibitem[Sridhara et~al\mbox{.}(2010)]%
        {sridhara2010towards}
\bibfield{author}{\bibinfo{person}{Giriprasad Sridhara}, \bibinfo{person}{Emily Hill}, \bibinfo{person}{Divya Muppaneni}, \bibinfo{person}{Lori Pollock}, {and} \bibinfo{person}{K Vijay-Shanker}.} \bibinfo{year}{2010}\natexlab{}.
\newblock \showarticletitle{Towards automatically generating summary comments for java methods}. In \bibinfo{booktitle}{\emph{Proceedings of the 25th IEEE/ACM international conference on Automated software engineering}}. \bibinfo{pages}{43--52}.
\newblock


\bibitem[Sridhara et~al\mbox{.}(2011a)]%
        {sridhara2011automatically}
\bibfield{author}{\bibinfo{person}{Giriprasad Sridhara}, \bibinfo{person}{Lori Pollock}, {and} \bibinfo{person}{K Vijay-Shanker}.} \bibinfo{year}{2011}\natexlab{a}.
\newblock \showarticletitle{Automatically detecting and describing high level actions within methods}. In \bibinfo{booktitle}{\emph{Proceedings of the 33rd International Conference on Software Engineering}}. \bibinfo{pages}{101--110}.
\newblock


\bibitem[Sridhara et~al\mbox{.}(2011b)]%
        {sridhara2011generating}
\bibfield{author}{\bibinfo{person}{Giriprasad Sridhara}, \bibinfo{person}{Lori Pollock}, {and} \bibinfo{person}{K Vijay-Shanker}.} \bibinfo{year}{2011}\natexlab{b}.
\newblock \showarticletitle{Generating parameter comments and integrating with method summaries}. In \bibinfo{booktitle}{\emph{2011 IEEE 19th international conference on program comprehension}}. IEEE, \bibinfo{pages}{71--80}.
\newblock


\bibitem[Steidl et~al\mbox{.}(2013)]%
        {6613836}
\bibfield{author}{\bibinfo{person}{Daniela Steidl}, \bibinfo{person}{Benjamin Hummel}, {and} \bibinfo{person}{Elmar Juergens}.} \bibinfo{year}{2013}\natexlab{}.
\newblock \showarticletitle{Quality analysis of source code comments}. In \bibinfo{booktitle}{\emph{2013 21st International Conference on Program Comprehension (ICPC)}}. \bibinfo{pages}{83--92}.
\newblock
\href{https://doi.org/10.1109/ICPC.2013.6613836}{doi:\nolinkurl{10.1109/ICPC.2013.6613836}}


\bibitem[Su and McMillan(2024)]%
        {su2024distilled}
\bibfield{author}{\bibinfo{person}{Chia-Yi Su} {and} \bibinfo{person}{Collin McMillan}.} \bibinfo{year}{2024}\natexlab{}.
\newblock \showarticletitle{Distilled GPT for source code summarization}.
\newblock \bibinfo{journal}{\emph{Automated Software Engineering}} \bibinfo{volume}{31}, \bibinfo{number}{1} (\bibinfo{year}{2024}), \bibinfo{pages}{22}.
\newblock


\bibitem[Sun et~al\mbox{.}(2024)]%
        {sun2024source}
\bibfield{author}{\bibinfo{person}{Weisong Sun}, \bibinfo{person}{Yun Miao}, \bibinfo{person}{Yuekang Li}, \bibinfo{person}{Hongyu Zhang}, \bibinfo{person}{Chunrong Fang}, \bibinfo{person}{Yi Liu}, \bibinfo{person}{Gelei Deng}, \bibinfo{person}{Yang Liu}, {and} \bibinfo{person}{Zhenyu Chen}.} \bibinfo{year}{2024}\natexlab{}.
\newblock \showarticletitle{Source code summarization in the era of large language models}.
\newblock \bibinfo{journal}{\emph{arXiv preprint arXiv:2407.07959}} (\bibinfo{year}{2024}).
\newblock


\bibitem[Tan et~al\mbox{.}(2024)]%
        {tan2024judgebench}
\bibfield{author}{\bibinfo{person}{Sijun Tan}, \bibinfo{person}{Siyuan Zhuang}, \bibinfo{person}{Kyle Montgomery}, \bibinfo{person}{William~Y Tang}, \bibinfo{person}{Alejandro Cuadron}, \bibinfo{person}{Chenguang Wang}, \bibinfo{person}{Raluca~Ada Popa}, {and} \bibinfo{person}{Ion Stoica}.} \bibinfo{year}{2024}\natexlab{}.
\newblock \showarticletitle{Judgebench: A benchmark for evaluating llm-based judges}.
\newblock \bibinfo{journal}{\emph{arXiv preprint arXiv:2410.12784}} (\bibinfo{year}{2024}).
\newblock


\bibitem[Taulli(2020)]%
        {taulli2020cobol}
\bibfield{author}{\bibinfo{person}{Tom Taulli}.} \bibinfo{year}{2020}\natexlab{}.
\newblock \showarticletitle{COBOL language: Call it a comeback?}
\newblock \bibinfo{journal}{\emph{Retrieved January}}  \bibinfo{volume}{13} (\bibinfo{year}{2020}), \bibinfo{pages}{2022}.
\newblock


\bibitem[Team et~al\mbox{.}(2024)]%
        {team2024gemini}
\bibfield{author}{\bibinfo{person}{Gemini Team}, \bibinfo{person}{Petko Georgiev}, \bibinfo{person}{Ving~Ian Lei}, \bibinfo{person}{Ryan Burnell}, \bibinfo{person}{Libin Bai}, \bibinfo{person}{Anmol Gulati}, \bibinfo{person}{Garrett Tanzer}, \bibinfo{person}{Damien Vincent}, \bibinfo{person}{Zhufeng Pan}, \bibinfo{person}{Shibo Wang}, {et~al\mbox{.}}} \bibinfo{year}{2024}\natexlab{}.
\newblock \showarticletitle{Gemini 1.5: Unlocking multimodal understanding across millions of tokens of context}.
\newblock \bibinfo{journal}{\emph{arXiv preprint arXiv:2403.05530}} (\bibinfo{year}{2024}).
\newblock


\bibitem[Upadhaya(2023)]%
        {upadhaya2023understanding}
\bibfield{author}{\bibinfo{person}{Ashish Upadhaya}.} \bibinfo{year}{2023}\natexlab{}.
\newblock \showarticletitle{Understanding Legacy Software: The Current Relevance of COBOL}.
\newblock  (\bibinfo{year}{2023}).
\newblock


\bibitem[Venigalla and Chimalakonda(2022)]%
        {venigalla2022empirical}
\bibfield{author}{\bibinfo{person}{Akhila Sri~Manasa Venigalla} {and} \bibinfo{person}{Sridhar Chimalakonda}.} \bibinfo{year}{2022}\natexlab{}.
\newblock \showarticletitle{An Empirical Study On Correlation between Readme Content and Project Popularity}.
\newblock \bibinfo{journal}{\emph{arXiv preprint arXiv:2206.10772}} (\bibinfo{year}{2022}).
\newblock


\bibitem[Wu et~al\mbox{.}(2021)]%
        {wu2021recursively}
\bibfield{author}{\bibinfo{person}{Jeff Wu}, \bibinfo{person}{Long Ouyang}, \bibinfo{person}{Daniel~M Ziegler}, \bibinfo{person}{Nisan Stiennon}, \bibinfo{person}{Ryan Lowe}, \bibinfo{person}{Jan Leike}, {and} \bibinfo{person}{Paul Christiano}.} \bibinfo{year}{2021}\natexlab{}.
\newblock \showarticletitle{Recursively summarizing books with human feedback}.
\newblock \bibinfo{journal}{\emph{arXiv preprint arXiv:2109.10862}} (\bibinfo{year}{2021}).
\newblock


\bibitem[Zhai et~al\mbox{.}(2016)]%
        {zhai2016automatic}
\bibfield{author}{\bibinfo{person}{Juan Zhai}, \bibinfo{person}{Jianjun Huang}, \bibinfo{person}{Shiqing Ma}, \bibinfo{person}{Xiangyu Zhang}, \bibinfo{person}{Lin Tan}, \bibinfo{person}{Jianhua Zhao}, {and} \bibinfo{person}{Feng Qin}.} \bibinfo{year}{2016}\natexlab{}.
\newblock \showarticletitle{Automatic model generation from documentation for Java API functions}. In \bibinfo{booktitle}{\emph{Proceedings of the 38th International Conference on Software Engineering}}. \bibinfo{pages}{380--391}.
\newblock


\bibitem[Zhu et~al\mbox{.}(2023)]%
        {zhu2023judgelm}
\bibfield{author}{\bibinfo{person}{Lianghui Zhu}, \bibinfo{person}{Xinggang Wang}, {and} \bibinfo{person}{Xinlong Wang}.} \bibinfo{year}{2023}\natexlab{}.
\newblock \showarticletitle{Judgelm: Fine-tuned large language models are scalable judges}.
\newblock \bibinfo{journal}{\emph{arXiv preprint arXiv:2310.17631}} (\bibinfo{year}{2023}).
\newblock


\end{thebibliography}

\appendix

\end{document}